**Linewidth of the Laser Optical Frequency Comb with Arbitrary Temporal Profile.**

**JACOB B KHURGIN[1], NATHAN HENRY[1], DAVID BURGHOFF[2], QING HU[2],**

[1]*Whiting School of Engineering, Johns Hopkins University, Baltimore, MD, 21218, USA*

[2]*Department of Electrical Engineering and Computer Science, Research Laboratory of Electronics, Massachusetts Institute of Technology, Cambridge, Massachusetts, 02139, USA*

**Abstract:** For many applications Optical Frequency Combs (OFCs) require a high degree of temporal coherence (narrow linewidth). Commonly OFCs are generated in nonlinear media from a monochromatic narrow linewidth laser sources or from a mode-locked laser pulses but in the all-important mid-infrared (MIR) and terahertz (THz) regions of spectrum OFCs can be generated intrinsically by the free-running quantum cascade lasers (QCLs) with high efficiency. These combs do not look like conventional OFCs as the phases of each mode are different and in temporal domain the OFC is a seemingly random combination of amplitude- and phase-modulated signals rather than a short pulse. Despite this "pseudo-randomness", the experimental evidence suggests that the linewidth of the QCL OFC is just as narrow as that of a QCL operating in the single mode. While universally acknowledged, this seemingly observation is not fully understood. In this work we rigorously prove this fact by deriving the expression for the Schawlow-Townes linewidth of QCL OFC and offer a transparent physical interpretation based on orthogonality of laser modes, indicating that despite their very different temporal profiles MIR and THz QCL OFCs are just as good for most applications as any other OFC.

**Introduction**

Optical Frequency Combs (OFCs) has been enjoying a healthy increase in interest in the last decade due to their applications in metrology, frequency standards, and spectroscopy[1,2] . The major difference between OFC and an arbitrary signal with a discrete periodic spectrum (such as produced by a multi-mode laser) is that in OFC the phases of all spectral components are locked and it assures narrow linewidth and stability (in the sense that all the frequencies remain equidistant in the long term) which is essential for all the existing and potential OFC applications. Following original work[1,2] the OFCs are routinely generated by mode-locked lasers[3-5], that are subsequently spectrally broadened in the nonlinear fiber. Nonlinear processes (self-phase modulation in temporal domain or four-wave mixing in spectral domain) require dispersion compensation and once it is attained a short soliton-like pulse is formed in the time domain. More recently, another strategy has been advancing – using a continuous-wave narrow linewidth laser source to generate OFC in a nonlinear micro-resonator[6-8] – once again a soliton being formed indicating that all the spectral components are locked with the same phase[9-11]. Obviously the linewidth of micro-resonator or nonlinear fiber-generated OFC can be essentially the same as that of the laser that pumps it since neither scheme has spontaneous emission.

While OFCs has been most successful in the near-IR region of the spectrum, from the point of view of spectroscopy it is mid-infrared (MIR) and Terahertz (THz) regions of the spectra that are more interesting because they contain information about vibrational and rotational structure of many organic and inorganic substances. In the absence of a direct laser pump sources one must revert to using optical parametric oscillators[9,10,12] or to down converting of near infrared OFC[13,14] which greatly increases the complexity, dimensions, and cost of MIR OFC sources.

Most recently, though, an entirely new method of generating MIR[15] and THz[16] OFCs based on free running quantum cascade lasers (QCL)[17] has been developed and dual comb spectroscopic measurements have been performed using these unconventional OFCs. What makes these OFCs "unconventional" is the fact that OFC regime in QCL does not require any additional intra-cavity phase locking mechanism but is achieved by means of the four wave-mixing in the fast (picosecond) saturable gain medium. This fast response time is a salient feature of intersubband transitions in semiconductor quantum wells and is unique to QCL and it is the reason that in time domain QCL frequency comb[15,16] looks nothing like a short pulse but is a predominately frequency-modulated (FM) signal. As explained in[18,19] the QCL active medium is a fast negative gain, which exactly opposite from the fast saturable absorbers that enable production of short pulses in mode-locked laser. The effect of the fast saturable gain is the opposite – it favors constant intensity output which can be either a single mode lasing or a FM signal. The single mode regime has higher threshold due to formation of spatial and spectral while the multimode FM regime mitigates the spatial and spectral hole burning and is therefore a regular operational regime of free running QCL in which group velocity dispersion (GVD) has been compensated[19,20]. More recent measurements have shown that the actual operating regime of THz and MIR QCL OFCs is more complicated than a simple FM[21,22] and have a significant intensity modulation on it as explained in[23,24]. While the signal is obviously periodic with a period of cavity round trip, within this interval it appears entirely "random" which[23] is the best way to mitigate spatial hole burning. Despite this,

QCL OFCs has been very successfully used in spectroscopy[25,26] which indicates that all the relevant parameters of QCL OFCs are comparable to those in other OFCs.

One of the most important parameters describing OFC is the linewidth of each spectral line in the comb. According to the original work of Schawlow and Townes[27] refined by Lax[28] the linewidth of a single mode laser is inversely proportional to the power $P_{out}$, or to the total number of photons in the cavity $N_p$. For a multimode laser operating on N modes one would then expect that the linewidth of each line would be inversely proportional to the number of photons in that mode, or $P_{out}/N$, i.e. N times wider. But it is experimentally well known that the linewidth of the mode locked lasers is comparable to the linewidth a single mode laser of equal average power since all the modes are locked into the same phase[29,30]. The linewidth and phase noise of mode locked lasers has been theoretically explored in a number of works[31-33] where it was assumed that all the modes are locked into the same phase since until recently that was the only practical way to lock all the phases using saturable absorber or an active intensity of phase modulator. But as QCL OFC with its mostly FM character appeared it became important to understand what kind of linewidth can be achieved in them. Measurements[34] have shown that the linewidth is indeed very narrow, comparable to the linewidth of a single-mode QCL operating at the same power. This narrowness has been explained[34] by introducing a concept of "supermode" but without detailed derivation of the amount of noise going into that mode and just assuming a Langevin spontaneous emission term As shown below this approach is correct but requires stronger theoretical foundation.

This foundation is provided in this work with a simple derivation of Schawlow-Townes (ST) linewidth for arbitrary laser OFC based on orthogonality of modes and the fact that the spontaneous noise amplitudes in each modes are uncorrelated. We also provide a simple intuitive and physically transparent picture of why only a small fraction of spontaneous noise actually affects the linewidth of the arbitrary OFC. Our conclusion is that despite their seemingly "random" temporary profiles OFCs produced by the lasers with short gain recovery times (such as QCLs) are every bit as good as OFCs produced by single mode lasers and microresonators.

**Derivation of the ST linewidth of the arbitrary OFC**

We consider the situation in which the steady phase locking of the modes has already taken place and the total field can be written as

$$E(t,z) = \sum_{n=1}^{N} E_n(t) e^{-j\omega_n t} a_n(z) = A(t) \sum_{n=1}^{N} f_n e^{-j\omega_n t} a_n(z) \qquad (1)$$

where $\omega_n$ is a frequency of n-th mode, $a_n(z) \sim \sin(\omega_n z / v)$ is the normalized shape of the n-th orthogonal mode, $v$ is the phase velocity of light, $\int a_m a_n dz = \delta_{mn}$, and the Fourier amplitudes $f_n$ have been normalized as $\sum_{n=1}^{N} |f_n|^2 = 1$. The average power can then be defined as $\bar{P} = |A|^2$. The mode amplitudes $E_n(t)$ as well as the total field amplitude $A(t)$ only experience the temporal change on the time scale that is much longer than the optical period $2\pi\omega_n^{-1}$. The normalized distribution of normalized instant electric field $E(z)$ inside the cavity and one of the modes $a_m(z)$ are shown in Fig.1a

Let us now write the expression for the temporal development of the slow-varying amplitude of the n-th mode[18,19]

$$\frac{dE_n(t)}{dt} = \frac{1}{2} g_n(\bar{P}) E_n - \frac{1}{2} \frac{E_n}{\tau_c} + i D_n E_n - \frac{E_n}{2\tau_c} + \sum_{l \neq n}^{N} \sum_{m \neq l, n}^{N} g_{nlm}(\bar{P}) \kappa_{nlm} E_l E_m^* E_{n+m-l} + S_n(t) \qquad (2)$$

where $\frac{1}{2} g_n(\bar{P})$ is the gain that includes both self-saturation and cross-saturation, $g_{nlm}(\bar{P})\kappa_{nlm}$ is the four-wave mixing term (but in case of active mode locking it may have a different shape corresponding to the sideband generation), $\tau_c$ is the cavity lifetime, $D_n = 2n^2\pi^2\tau_c^{-2}\beta_2 v_g$ is the dispersive term, $v_g$ is the velocity and $\beta_2$ is the GVD and $S_n(t)$ is the noise complex amplitude in the n-th mode. The noise can be caused by the spontaneous emission of photons in the n-th mode, but it can also be due to a far more mundane causes such as cavity length oscillations due to random vibrations. The noise amplitudes of different modes are not correlated, $\langle S_n^*(t) S_m(t') \rangle = \delta_{nm} \delta(t - t')$ Upon substituting (1) into (2) we obtain

$$\frac{dA(t)}{dt} f_n = \frac{1}{2} g_n(\bar{P}) A(t) f_n + i D_n A(t) f_n - \frac{A(t) f_n}{2\tau_c} + A(t)|A(t)|^2 \sum_{l \neq n}^{N} \sum_{m \neq l, n}^{N} g_{nlm}(\bar{P}) \kappa_{nlm} f_l f_m^* f_{n+m-l} + S_n(t)$$

(3)

The terms under double summation are the four-wave-mixing (FWM) terms that have "amplitude" or in-phase parts and "phase" or quadrature parts. The quadrature terms causes frequency chirp and once the stable regime has been reached (i.e. once the relative phases and amplitudes of modes $f_n$ get locked while the amplitude of the envelope $A(t)$ can still experience change on a slow scale) they get compensated by the group velocity dispersion terms $D_n$, i.e.

$$\text{Im} \sum_{l \neq n}^{N} \sum_{m \neq l,n}^{N} g_{nlm}(\bar{P}) \kappa_{nlm} f_l f_m^* f_{n+m-l} + i D_n A(t) f_n = 0 \tag{4}$$

Multiply now both sides of (3) by $f_n^*$ and perform summation. The amplitude terms only cause the energy redistribution between the modes hence once the stable operational regime has been reached they all sum up to zero. i.e.

$$\text{Re} \sum_{n}^{N} \sum_{l \neq n}^{N} \sum_{m \neq l,n}^{N} g_{nlm}(\bar{P}) \kappa_{nlm} f_l f_m^* f_{n+m-l} f_n^* = 0 \tag{5}$$

The new differential equation then becomes

$$\frac{dA(t)}{dt} = \frac{1}{2} g(\bar{P}) A(t) - \frac{A(t)}{2\tau_c} + \sum_{n=1}^{N} S_n(t) f_n^* \tag{6}$$

Note that while we have mentioned that the steady state regime has been obtained the temporary dependence of the amplitude is still present in (6) to describe the changes in the complex amplitude due to relatively small variations in pump power that do not change the phase relations between modes. Obviously any small deviations from (4) and (5) can be treated as additional sources of noise and simply added to (6). Let us now examine the noise term. Since the phases of complex noise amplitudes in each mode are random one must sum up the noise powers, i.e. total noise power is $|S_{tot}|^2 = \sum_{n=1}^{N} |S_n|^2 |f_n|^2$.

Under a reasonable assumption that all the noise powers are identical, i.e. $|S_n| \equiv |S|$ we obtain $|S_{tot}|^2 = |S|^2$, i.e. the total noise added to a given stable phase-locked signal is equal to the noise in any given mode, and we have

$$\frac{dA(t)}{dt} = \frac{1}{2} g(\bar{P}) A(t) - \frac{A(t)}{2\tau_c} + S(t) \tag{7}$$

In other words, this equation looks exactly like the equation for a single mode laser and is therefore bound to yield the linewidth that is determined by the total average power $\bar{P} = |A|^2$ rather than by a power in a given mode. One easy way to interpret this situation is to simply assume that one can take any linear combination of modes in the cavity and just call it an eigenmode of the system[34].

Of course, as different modes do have different frequencies their linear combination does not necessarily amount to an eigenmode, which after all is supposed to have a well-defined frequency, but this is still a useful analogy which allows equation (7) to be obtained instantly.

One can now obtain the ST linewidth for the case when the only noise source is the spontaneous emission by re-writing (7) by introducing the photon number in the cavity as $N_p = |A|^2 \tau_{rt} / \hbar\omega$ where $\tau_{rt}$ is the cavity round trip time, so that

$$\frac{dN_p}{dt} = \left[g(N_p) - \frac{1}{\tau_c}\right] N_p + n_{sp} g(N_p) \tag{8}$$

where $n_{sp} \geq 1$ is the spontaneous emission factor caused by the incomplete population inversion. In steady state the saturated gain is $g(N_p) = N_p \tau_c^{-1} / (N_p + n_{sp})$ and the linewidth can be introduced as the effective photon decay rate,

$$\Delta\nu_{ST} = [\tau_c^{-1} - g(N_p)]/2\pi = \frac{n_{sp}}{N_p + n_{sp}} \tau_c^{-1} / 2\pi \approx \Delta\nu_0 n_{sp} N_p^{-1} \tag{9}$$

where $\Delta\nu_0 = \tau_c^{-1}/2\pi$ is the cold cavity linewidth. Relating the photon density inside the cavity to the output power as $P_{out} = \hbar\omega \eta_{out} \tau_c^{-1} N_p$ where $\eta_{out}$ is the outcoupling efficiency, the ST linewidth assumes a more familiar shape $\Delta\nu_{ST} = 2\pi\eta_{out}\Delta\nu_0^2 \hbar\omega / P_{out}$

**Physical interpretation**

Let us now consider the physical interpretation of the result – how come that despite the fact that each of N modes contributes to the spontaneous emission, only $1/N_{th}$ of that radiation contributes to both phase and amplitude noise of the laser comb? For a short (mode-locked with all phases equal) pulse the interpretation in time domain is obvious: the pulse sequence containing N cavity modes has the duty cycle of exactly 1/N – hence only 1 of each N spontaneously emitted photons will affect the phase (and amplitude) of the pulse – the rest of them will appear as additive noise which can always be filtered out. For FM frequency comb a similar time domain interpretation can be made: assume for simplicity that the laser simply jumps from one mode to another – i.e. instant frequency is always equal to one of the cavity frequencies. Obviously the spontaneous emission into other N-1 modes does not contribute to the multiplicative phase noise but simply presents an additive noise.

To understand the narrow linewidth of the arbitrary comb rather than using temporal domain it is helpful to turn instead to the spatial dependence of the signal (1) and consider the change of phase imposed by the emission of a single photon into the m-th mode. According to (8) each photon is emitted every $\delta t_m = \tau_c / n_{sp}$ and the power of that photon, $\delta P_m = |\delta E_m|^2 = \hbar\omega\tau_{rt}^{-1}$. Therefore, the phase change imposed by this photon is (see Fig.1a)

$$\delta\varphi_m = \int E(t,z)\delta E_{mQ} a_m(z)dz / \int E^2(t,z)dz = f_m \delta E_{mQ} / |A| \qquad (10)$$

where $\delta E_{mQ}$ is the quadrature component of the spontaneous emission and we have used the orthogonality of the laser modes in space. Now, in the time interval $\Delta t$ there will be $\Delta N_m = \Delta t / \delta t_m = n_{sp}\tau_c^{-1}\Delta t$ photons emitted into the m-th mode and according to random walk process theory the variance of the phase will be

$$\langle\delta\varphi^2\rangle = |f_m|^2 \frac{|\delta E_{mQ}|^2}{|A|^2}\Delta N_m = \frac{\hbar\omega n_{sp}\tau_c^{-1}|f_m|^2}{|A|^2}\frac{\Delta t}{\tau_{rt}} = \frac{n_{sp}\tau_c^{-1}}{N_p}|f_m|^2 \Delta t \qquad (11)$$

But according to the theory for the random walk process with the variance $\langle\delta\varphi^2\rangle = 2\pi C \Delta t$ the power spectral density of the phase noise can be found as $S_\varphi(\omega) = C/\omega^2$ with the linewidth being $\Delta\nu = C$. Therefore, the linewidth can be found as

$$\Delta\nu = \langle\delta\varphi^2\rangle / 2\pi\Delta t = \frac{n_{sp}\Delta\nu_0}{N_p}|f_m|^2 \qquad (12)$$

Comparing (12) with (9) one can state that the linewidth's contribution from spontaneous emission in the m-th mode is proportional to its relative weight $f_m^2$, i.e. roughly 1/N. Thus physical interpretation in spatial domain is simple – since the spontaneous emission into a given resonator mode only weakly overlaps in space with actual distribution of the laser field inside the cavity, most of this emission ends up as an additive noise and does not contribute to the phase/frequency noise and linewidth. This situation is schematically explained in Fig.1b

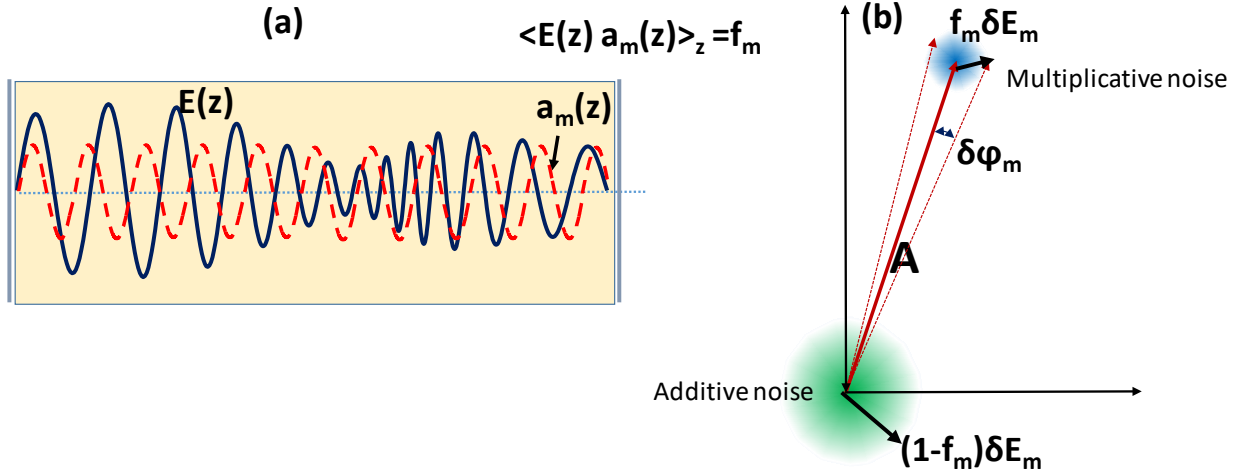

**Figure 1.** (a) spatial distribution of one of the cavity modes $a_m(z)$ and instant field $E(z)$ inside the laser cavity (b) Phasor representation of the phase noise and linewidth in arbitrary OFC. The amplitude of the coherent electric field OFC is $A$. The electric field of a photon emitted into the m-th mode is $\delta E_m$. The averaged over the cavity length product of the two fields, proportional to the weight of the m-th mode in the total laser field $f_m$, engenders the multiplicative noise that in the end contributes to the linewidth. The rest of the spontaneous electric field, proportional to $1-f_m$ is orthogonal in space to the coherent field and only generates additive noise.

## Discussion and Conclusions

The key conclusion reached in this work is that we have rigorously confirmed the previously made argument that the linewidth of a given multi-mode phase locked laser emission does not depend on exactly what is the phase relation between the individual modes as long as this relation exits, i.e. as long as the locking mechanism (which can be a fast saturable absorption as in conventional mode locked laser or a fast saturable gain in QCL) is strong enough to overcome dispersion of group velocity and gain. The OFC signal may indeed be a short pulse[1], an FM signal[18], or a combined AM/FM signal[21,22] - from the point of view of most practical OFC applications the character of comb in time domain is irrelevant and the measurements with all kinds of combs can be expected to yield the results with roughly the same precision.

Let us compare OFC obtained directly from the phase-locked multimode laser with the one generated in a micro-resonator pumped by a single mode laser[10]. For equal average powers both lasers would have similar linewidth but the multimode laser will have substantially higher additive noise due to spontaneous emission. This noise can easily be filtered out from the single mode laser.

But the efficiency of the laser OFC generation is typically much higher (precisely due to the presence of laser gain at each mode) therefore the micro-resonator comb must be amplified and the spontaneous emission in the amplifier will of course generate significant additive noise which will negate all the purported advantages of micro-resonator comb.

Therefore, an OFC generated by the free-running lasers with fast saturable gain media, such as QCLs can provide measurements just as good if not superior to the comb produced using nonlinear processes in microresonators given comparable signal to noise ratios combined the simplicity, smaller size and higher efficiency of the free running laser OFCs

**Acknowledgement:** The authors acknowledge generous support provided by DARPA SCOUT program. Additionally, they would like to thank Prof. Jérôme Faist and Matt Singleton at ETH Zurich for many stimulating discussions.